\documentclass[10pt,conference,a4paper,twocolumn] {IEEEtran}

\usepackage{graphicx}
\usepackage{epic}
\usepackage{figsize}
\usepackage{float}
\usepackage{hyperref}
\usepackage[section]{placeins}
\usepackage{cite}
\usepackage{amsmath}
\usepackage{amssymb}
\usepackage{amsfonts}

\usepackage[utf8]{inputenc}
\usepackage{cite}
\usepackage{amsmath,amssymb,amsfonts}
\usepackage{graphicx}
\usepackage{textcomp}
\usepackage{xcolor}
\usepackage{algorithm} 
\usepackage{algpseudocode}
\usepackage{array}
\usepackage{epic}
\usepackage{figsize}
\usepackage{float}
\usepackage{hyperref}
\usepackage[section]{placeins}

\newcommand{\email}[1]{{\fontfamily{cmtt}\selectfont#1\normalfont}}

\IEEEoverridecommandlockouts

\title{Adversarial Deep Reinforcement Learning for Cyber Security in Software Defined Networks}

\author{\IEEEauthorblockN{Luke Borchjes\IEEEauthorrefmark{1},Clement Nyirenda\IEEEauthorrefmark{2},Louise Leenen\IEEEauthorrefmark{3}}
\IEEEauthorblockA{
    \emph{Computer Science Department, University of the Western Cape}\\ 
    \emph{Robert Sobukwe Rd, Bellville, 7535, South Africa}\\
$^1$\email{ldborchjes@gmail.com, 3647745@myuwc.ac.za}\\
$^2$\email{cnyirenda@uwc.ac.za}\\
$^3$\email{lleenen@uwc.ac.za}\\
}}
\begin{document}
\maketitle
\begin{abstract}
This paper focuses on the impact of leveraging autonomous offensive approaches in Deep Reinforcement Learning (DRL) to train more robust agents by exploring the impact of applying adversarial learning to DRL for autonomous security in Software Defined Networks (SDN). Two algorithms, Double Deep Q-Networks (DDQN) and Neural Episodic Control to Deep Q-Network (NEC2DQN or N2D), are compared. NEC2DQN was proposed in 2018 and is a new member of the deep q-network (DQN) family of algorithms. The attacker has full observability of the environment and access to a causative attack that uses state manipulation in an attempt to poison the learning process. The implementation of the attack is done under a white-box setting, in which the attacker has access to the defender's model and experiences. Two games are played; in the first game, DDQN is a defender and N2D is an attacker, and in second game, the roles are reversed. The games are played twice; first, without an active causative attack and secondly, with an active causative attack. For execution, three sets of game results are recorded in which a single set consists of 10 game runs. The before and after results are then compared in order to see if there was actually an improvement or degradation.  The results show that with minute parameter changes made to the algorithms, there was growth in the attacker's role, since it is able to win games. Implementation of the adversarial learning by the introduction of the causative attack showed the algorithms are still able to defend the network according to their strengths.
\end{abstract}

\vspace{0.2cm}

\section{Introduction}

Software Defined Networking (SDN) is a three-layer network architecture that has been in practice since 2013 \cite{bailey2013sdn}. Comprising an application layer, a control layer, and an infrastructure layer, SDN delivers a robust framework for managing network tasks \cite{Yi_Han_et._al.}. One of the major advantages of SDN is its separation of network control and forwarding functions, which enables the controller to be programmed for various application services and tasks. This separation facilitates the convenient management, configuration, and optimisation of network resources using standardised protocols. It is also shown in \cite{xie2018survey} that machine learning has many uses in SDN. The COVID-19 pandemic led to a significant increase in telecom users, driving further investment in SDN, an essential technology for realising the potential of 5G networks. These 5G networks are expected to bring substantial improvements to the telecom industry \cite{syed2020software}. However, this surge has also led to an increase in cybercrime \cite{baz2021impact, govender2021global}, highlighting the ever-present need for enhanced network security.

This work focuses on employing adversarial learning in the training and implementation of model-free deep reinforcement learning in Software-Defined Networking (SDN). The rise of AI models and algorithms has been significant but met with increased scepticism. Despite this, the automation capabilities of SDN have positioned it as a strong candidate for autonomous defence mechanisms \cite{Yi_Han_et._al., Yi_Han_et._al.2}, prompting its broad adoption across various industries. Achieving robustness in AI model implementation has proven to be crucial since attackers are perpetually attempting to exploit vulnerabilities in the learning process. Thus, it is essential to cultivate models capable of tolerating corrupted or malicious inputs, the work done in \cite{tang2016deep} emphasizes the importance of this. In this regard, the attacking agent utilises a data poisoning attack, implemented through state manipulation \cite{Yi_Han_et._al.2}. Experiences used for training are manipulated by implanting false positives and negatives, building on the previous work cited \cite{b1}, wherein two model-free deep reinforcement learning algorithms, double deep q-learning and neural episodic control to deep q-network, were juxtaposed. In \cite{b1}, they were implemented into a software defined network running a capture the flag (CTF) game. The game was set up such that one agent had to defend the network against the other with the goal to measure and compare performances. The same game setting is used in this experimentation. 

The remainder of this paper is summarised as follows: Section \ref{DRL4SDN} summarises the problem faced in previous work when applying the model-free deep reinforcement learning agents to a software defined networks; Section \ref{env} introduces the environment employed in the investigation of the work, which is kept the same as in previous work and inspired material; Section \ref{results} covers the results of the investigation, in which the win rates and performance is evaluated; Section \ref{conclusion} concludes the investigation.

\section{Problem: Deep Reinforcement Learning for Cybersecurity in Software Defined Networking}\label{DRL4SDN}

\subsection{Background on Software Defined Networking}\label{BSDN}

Software-defined networking (SDN) is an approach to network management that allows dynamic, efficient network configuration in order to improve network performance \cite{bailey2013sdn}. It was spawned as the result of the desire to separate the data plane from the control plane \cite{bailey2013sdn}. SDN is composed of the three layers: (1) application layer; (2) Control layer; (3) infrastructure layer. The application layer is made up of applications which deliver services and communicate their network requirements to the controller using northbound APIs. The Control layer hosts the SDN controller, translates requirements into low-level controls that are then sent to the infrastructure layer using southbound API's. The infrastructure layer consists of network switches and other infrastructural components \cite{bailey2013sdn, Yi_Han_et._al.}. The major advantage of SDN is that it separates network control and forwarding functions, allowing the controller to be programmable to perform various application services and tasks \cite{bailey2013sdn, Yi_Han_et._al.}. Consequently, network resources can be conveniently managed, configured and optimised using the standardised protocols. Due to its architecture there has been a good variety of available open-source SDN controller platforms/frameworks, a few examples being OpenDayLight, RYU, NOX/POX and Open vSwitch \cite{bailey2013sdn, Yi_Han_et._al.}.

\subsection{Background on Reinforcement Learning}\label{BRL}

Reinforcement Learning (RL) deals with a sequential decision making problem where an agent interacts with the environment to maximise its rewards, implemented as a Markov Decision Process (MDP). An MDP is classified as follows: $(S, A, P, R, GAMMA (\gamma))$ \cite{Otterlo2012MArkovDP} where each time step $t$, the agent (1) receives an observation $s_t$ ($S$) of the environment; (2) takes an action $a_t$ ($A$) based on its policy $\pi$ ($P$), which is a mapping from states to actions; and (3) obtains a reward $r_t$($R$) based on state $s_t$, action $a_t$, and the environment’s transition to a new state $s_{t+1}$. The goal of the agent is to maximise its cumulative rewards, i.e., $R_t$ = $\sum_{\tau=t}^{\infty} \gamma^{\tau-t}r_{\tau}$, where $\gamma \in (0,1]$ is a discount factor which affects the present importance of long-term rewards \cite{Yi_Han_et._al.2}. The focus of experimentation was on a well known Deep RL algorithm — Double Deep Q-Networks (DDQN) \cite{Van_Hasselt_Hado_2010} and new variant Neural Episodic Control to Deep Q-Network (NEC2DQN) \cite{nishio2018faster} — and their ability to perform.

\paragraph{Double Deep Q-Learning}

To solve the overestimation of action-values the algorithm Double Q-Learning is proposed. Double Q-Learning is the implementation of two Q functions: $Q_A$ and $Q_B$. Each Q function is updated from the other’s next state \cite{Van_Hasselt_Hado_2010}. Its creation was the result of combating the over estimation problem well known with DQL as the maximisation bias \cite{Van_Hasselt_Hado_2010}.

\paragraph{Neural Episodic Control to Deep Q-Network}\label{NEC2DQN}
Neural Episodic Control (NEC), proposed in \cite{Alexander_Pritzel_et_al_2017}, can execute successful strategies as soon as they are experienced, instead of waiting for optimization mechanisms, such as stochastic gradient descent, to be done as is the case with DQN. Nevertheless, NEC becomes very memory intensive in latter stages; this is where a DQN is introduced, since both converge to a $Q$ value. A DQN can be trained from NEC and once a certain point of convergence is reached, the load can be shifted from the NEC to DQN. The shift from one to the other is gradual, but at a point, the change step $CS$, NEC is no longer used for decision making, but only training and evaluation, and decision making is done using the DQN. In this paper we decided to make the $CS$ occur after the first 20\% of turns have been passed.  

\subsection{Model-Free Deep Reinforcement Learning for Cybersecurity in Software Defined Networking}\label{MFDRL}

In \cite{b1}, where DDQN and N2D were implemented, tested and compared for cybersecurity within an SDN framework, the goal was to investigate using deep reinforcement learning for autonomous network defence. DDQN, a well known and matured algorithm was placed against a relatively newer algorithm N2D. N2D was chosen because it was designed to overcome the limitations of both NEC and DQN and has been shown to perform better than DDQN in certain cases \cite{nishio2018faster}. Two-tailed t-test analysis of results was done to determine if one was better than the other, by determining if there was any statistical difference, however the results showed that there was none. Therefore, DDQN was determined to be the more favourable due to its simplicity. The work also served as a baseline of what can be expected as well as have a reference point to reflect on when analysing newer results from changes. 

While the work cited previously showed promise, there were notable limitations and concerns \cite{b1}. One significant issue was the defender's domination of all game runs. On the surface, this bias towards the preferred outcome seems beneficial, but a deeper look reveals room for improvement, particularly from an attacker's perspective. More balanced engagement between the players would foster better learning for both agents, mitigating the environment's apparent defender bias. To counteract this bias, the attacker was permitted full observability of the environment.

The work in \cite{b1} proposed increasing the number of game runs to offer more total steps for each agent and a larger data pool for analysis. Furthermore, it suggested the implementation of adversarial learning, with the attacker conducting a causative attack on the defending agent, alongside the expansion of the network topology. It's important to note that while improving these algorithms may yield diminishing returns as a defender, there could be significant growth as an attacker. The game environment's inherent bias towards the defender means improving these algorithms may also make them more effective as offensive tools within the cybersecurity space, highlighting the potential for growth in the attacking role.

\subsection{Adversarial Machine Learning}\label{SSAML}

Adversarial machine learning is the study of the attacks on machine learning algorithms and is used in machine learning to misguide a model with malicious input \cite{kianpour}. It has also been shown that by maliciously altering the input for Deep Neural Networks with adversarial attacks it can easily be fooled into predicting the wrong label \cite{goodfellow}. The purpose of adversarial machine learning is not to emphasise the flaws of these algorithms, but to leverage these attacks during training as a means of training more robust agents \cite{apattanaik}. Most deployed cyber defence solutions are still rule-based and require human involvement, this opens the opportunity for false alarms \cite{Yi_Han_et._al.2}. Training robust agents through adversarial learning could help against any possible false alarms, allowing them to still make optimal decisions. 

In this investigation a data poisoning attack was chosen and done by the perturbation of the input for the agents. Since a state $s_{t}$ at any step $t$ is an array of length 80 containing binary digits $d \in [0, 1]$, the attack was implemented as the injection of false positives $(FPs)$ and false negatives $(FNs)$. The original observed experience is $(s, a, s', r)$ but instead the agent observes the new tampered experience $(s, a, s'+\delta, r')$ instead.  The implementation of the adversarial learning attack is described in section \ref{env}.

\section{Environment}\label{env}

\begin{algorithm}[h]
    \caption{State manipulation attack originally from \cite{Yi_Han_et._al.}.}
    \label{atk_algorithm}
    \begin{algorithmic}[1]
        \State \textbf{INPUT:} Original experience $(s, a, s', r)$
        \State Limit on number of FPs and FNs: $LIMIT$
        \State \textbf{OUTPUT:} Original experience $(s, a, s'+\delta, r')$
        \State $FP$ = $FN$ = $[ \hspace{0.1 cm} ]$
        \State $minQ_{FP} = minQ_{FN} = [ \hspace{0.1 cm} ] $
        \State 
        \For {node in State}
            \If {$ node$ is uncompromised mark as compromised }
                \If {$Q(s'+\delta, a') < 1$ or $Q(s'+\delta, a') <$ any value in $minQ_{FN}$}
                    \State Insert $FN$ into $FN$ and $minQ_{FN}$ 
                    \If{$|FN| < LIMIT$}
                        \State remove extra nodes from $FN$ and $minQ_{FN}$
                    \EndIf
                \EndIf
                \State restore $node$ as uncompromised  
            \EndIf
            \If {$node$ is compromised mark as uncompromised}
                \If {$Q(s'+\delta, a') < 1$ or $Q(s'+\delta, a') <$ any value in $minQ_{FP}$}
                    \State Insert $FP$ into $FP$ and $minQ_{FP}$ 
                    \If{$|FP| < LIMIT$}
                        \State Remove extra nodes from $FP$ and $minQ_{FP}$
                    \EndIf
                \EndIf
                \State restore $node$ as compromised
            \EndIf
        \EndFor
        \State Change nodes in $FN$ to uncompromised
        \State Change nodes in $FP$ to compromised
        \State \textbf{return} $(s, a, s'+\delta, r')$
    \end{algorithmic}
\end{algorithm}

The adversarial machine learning attack implemented in this research is a state manipulation attack, which was adopted from \cite{Yi_Han_et._al.}, and is presented in algorithm \ref{atk_algorithm}. In \cite{Yi_Han_et._al.}, two adversarial attacks were implemented; the first being the flipping of reward signs, and the second being a data poisoning attack done through state perturbation (manipulation). However, in \cite{Yi_Han_et._al.2}, they stated that the flipping reward sign attack proved to have made little to no impact; therefore in this work we have chosen to omit it and focus solely on the state manipulation attack.

As mentioned in section \ref{DRL4SDN} subsection \ref{SSAML}, the experience of the defending agent is poisoned by the injection of false positives and false negatives in the state. Slight changes were made from the original in \cite{Yi_Han_et._al.}, the core however remains the same. In our case we input the original state and loop over the part of the state that contains the nodes.

Our environment utilised an SDN network, composed of four subnets with a total of 32 hosts and 48 visible links, integrated with a CTF game \cite{b1}. Just as in previous research, three starting points for the attacker were chosen, and a critical server flag was established as the attacker's goal \cite{b1}. The attacker targets the training step of the defending agent and operates under a white-box setting, where the attacker has direct access to the experiences and model of the defender \cite{Yi_Han_et._al.}. If a black-box setting had been chosen, the attacker would need to train a surrogate model and select the appropriate nodes to falsify based on that surrogate model \cite{Yi_Han_et._al.2}.

In the games, the players are the attacking and the defending agents. Games are categorised according to which agent is attacking and defending. For game 1, the attacker is the agent using DDQN and the defender is the agent using N2D. For game 2 these roles are reversed. Subsequently each game is played initially without the attack, this means that no adversarial learning takes place. The games are then played again with the inclusion of the attack, introducing adversarial learning. The CTF game was implemented in the same SDN emulation used in \cite{b1} in which the SDN was built using MiniNet with RYU as the network controller of choice. A star topology was used for the SDN, with four subnets. Subnet 1 contains 6 hosts, subnet 2 contains 8 hosts, subnet 3 contains 9 hosts, and subnet four contains 9 hosts.

\section{Results}\label{results}
The following results are representative of the performance of the agents in their roles. In this work we take the results of multiple different game sets. For each game we have 3 sets, each containing 10 consecutive game runs. Set 1 contains games played with 5,000 turns, set 2 contains games played with 50,000, set 3 contains games played with 500,000 turns. Game sets of multiple turn counts are recorded due to the change step functionality of the NEC2DQN algorithm, thus the algorithm will function differently according to the amount of total steps in the game. Having results over three sets allows us to see the impact of a varying change step value, as mentioned in section \ref{DRL4SDN} subsection \ref{BRL}. In addition we also get to see the scaling of DDQN since DDQN is set to have a better performance in longer games. 

These results are discussed according to their game, afterwhich their results are analysed and their implications to SDN are discussed. It should be noted that multiple outcomes could occur as a result of the inclusion of the attack but the following is considered; (1) An agent could win more games but with an increase in average amount of turns, (2) An agent can win less games but have an improved turn count, (3) the agent could win more with an improved turn count and (4) there could be no change at all. Only outcomes 1 and 3 are confident indicators of improvement, 2 however is more subjective to the situation.

\begin{enumerate}
    \item \textbf{\textit{Game 1 Results:}}\label{} Table \ref{table_1} shows the control results for the different sets of game 1 without the causative attack active. The attacking agent uses DDQN and the defending agent uses N2D. For set 1 the results are 7 - 3 in favour of the defender. The attacking agent managed to win games 1, 6 and 10. The defending agent on the hand managed to hold back the defender for the entirety of the game's duration. The defender won all games by means of outlasting the attacker giving it an average of 5,000 turns, however, the attacker took on average 4,140 turns to win. For set 2 the results are 6 - 4 in favour of the defender. The attacker manages to win games 1, 3, 5 and 6. The defender however in the remainder of the set manages to isolate/remove the attacker from the network. On average it took the defender 7,401 number of turns to win and 5,611 for the attacker to win. For set 3 the results are 7 - 3 in favour of the attacker. Most of the runs in the game set are won by the attacker, with the exception of runs 5, 7, and 10. On average it took the defender 9,534 number of turns to win and 5,698 for the attacker to win.

    \begin{table}[htbp]
        \caption{Control for game 1}
        \begin{center}
        \begin{tabular}{|c|c|c|c|c|c|c|}
        \hline
        \textbf{Game}&\multicolumn{6}{|c|}{\textbf{Game sets}} \\
        \cline{2-7} 
        \textbf{Runs} & \multicolumn{2}{|c|}{\textbf{\textit{Set 1}}}& \multicolumn{2}{|c|}{\textbf{\textit{Set 2}}}& \multicolumn{2}{|c|}{\textbf{\textit{Set 3}}} \\
        \cline{2-7}
        \textbf{ } & \textbf{\textit{Agent}}&\textbf{\textit{Turn}}& \textbf{\textit{Agent}}&\textbf{\textit{Turn}}& \textbf{\textit{Agent}}&\textbf{\textit{Turn}}\\
        \hline
        1 & DDQN & 4914 & DDQN & 8034 & DDQN & 78 \\
        2 & N2D	& 5000 & N2D & 7785 & DDQN & 24 \\
        3 & N2D	& 5000 & DDQN & 2785 & DDQN & 9084 \\
        4 & N2D	& 5000 & N2D & 9174 & DDQN & 752 \\
        5 & N2D	& 5000 & DDQN & 4722 & N2D & 3051 \\
        6 & DDQN & 2670 & DDQN & 6904 & DDQN & 6750 \\
        7 & N2D	& 5000 & N2D & 2907 & N2D & 7783 \\
        8 & N2D	& 5000 & N2D & 2997 & DDQN & 10906 \\
        9 & DDQN & 4836 & N2D & 20575 & DDQN & 12294 \\
        10 & N2D & 5000 & N2D	& 969 & N2D	& 17769 \\
        \hline
        \multicolumn{7}{l}{$^{\mathrm{*}}$In this table DDQN is the attacker, N2D is the defender.}
        \end{tabular}
        \label{table_1}
        \end{center}
    \end{table}
    
    
    Table \ref{table_2} shows the game results for the different sets of game 1 with the causative attack active. For set 1 the results are 8 - 2 in favour of the defender, with only runs 1 and 2 being won by the attacker, and the latter being won by the defender. On average it took the defender 4,845 turns to win and 589 for the attacker to win. For set 2 the results are 6 - 4 in favour of the defender, with only runs 1 - 6 being won by the defender and the latter the attacker. On average it took the defender 9,800 turns to win and 7,625 turns for the attacker to win. For set 3 the results are 7 - 3 in favour of the attacker, with only runs 6, 8 and 9 being won by the defender. On average it took the defender 15,593 turns to win and 3,827 turns for the attacker to win.
    
    \begin{table}[htbp]
        \caption{After implementing causative attack in game 1}
        \begin{center}
        \begin{tabular}{|c|c|c|c|c|c|c|}
        \hline
        \textbf{Game}&\multicolumn{6}{|c|}{\textbf{Game sets}} \\
        \cline{2-7} 
        \textbf{Runs} & \multicolumn{2}{|c|}{\textbf{\textit{Set 1}}}& \multicolumn{2}{|c|}{\textbf{\textit{Set 2}}}& \multicolumn{2}{|c|}{\textbf{\textit{Set 3}}} \\
        \cline{2-7}
        \textbf{ } & \textbf{\textit{Agent}}&\textbf{\textit{Turn}}& \textbf{\textit{Agent}}&\textbf{\textit{Turn}}& \textbf{\textit{Agent}}&\textbf{\textit{Turn}}\\
        \hline
        1 & DDQN & 40 & N2D & 7733 & DDQN & 2900 \\
        2 & DDQN & 1138 & N2D & 8701 & DDQN & 1780 \\
        3 & N2D	& 99 & N2D & 3609 & DDQN & 4218 \\
        4 & N2D	& 4537 & N2D & 16511 & DDQN & 2386 \\
        5 & N2D	& 4281 & N2D & 3057 & DDQN & 7612 \\
        6 & N2D & 5000 & N2D & 19193 & N2D & 39033 \\
        7 & N2D	& 5000 & DDQN & 5978 & DDQN & 2076 \\
        8 & N2D	& 5000 & DDQN & 4736 & N2D & 4917 \\
        9 & N2D & 5000 & DDQN & 7572 & N2D & 2831 \\
        10 & N2D & 5000 & DDQN	& 12216 & DDQN & 5820 \\
        \hline
        \multicolumn{7}{l}{$^{\mathrm{*}}$In this table DDQN is the attacker, N2D is the defender.}
        \end{tabular}
        \label{table_2}
        \end{center}
    \end{table}

    \item \textbf{\textit{Game 2 Results:}} Table \ref{table_3} shows the control results for the different sets of game 2 without the causative attack active. The attacking agent is using N2D and the defending agent is using DDQN. For set 1 the results are 7 - 3 in favour of the defender. The attacking agent managed to win games 2, 5 and 8. The defending agent on the hand managed to hold back the defender for the entirety of the game's duration. On average it took the defender 5,000 turns to win and 1,632 turns for the attacker to win. For set 2 the results are 7 - 3 in favour of the attacker. The defender only manages to win games 1, 2 and 6. The defender however in the remainder of the set manages to isolate/remove the attacker from the network. On average it took the defender 907 turns to win and 2,406 turns for the attacker to win. For set 3 the results are 5 - 5 being an even split. The first half of the set is mostly dominated by the defender but the latter is the attacker. On average it took the defender 6,917 turns to win and 2,454 turns for the attacker to win.
    
    \begin{table}[htbp]
        \caption{Control for game 2}
        \begin{center}
        \begin{tabular}{|c|c|c|c|c|c|c|}
        \hline
        \textbf{Game}&\multicolumn{6}{|c|}{\textbf{Game sets}} \\
        \cline{2-7} 
        \textbf{Runs} & \multicolumn{2}{|c|}{\textbf{\textit{Set 1}}}& \multicolumn{2}{|c|}{\textbf{\textit{Set 2}}}& \multicolumn{2}{|c|}{\textbf{\textit{Set 3}}} \\
        \cline{2-7}
        \textbf{ } & \textbf{\textit{Agent}}&\textbf{\textit{Turn}}& \textbf{\textit{Agent}}&\textbf{\textit{Turn}}& \textbf{\textit{Agent}}&\textbf{\textit{Turn}}\\
        \hline
        1 & DDQN & 5000 & DDQN & 131 & DDQN & 81 \\
        2 & N2D & 1606 & DDQN & 1032 & DDQN & 1671 \\
        3 & DDQN & 5000 & N2D & 5825 & N2D & 570 \\
        4 & DDQN & 5000 & N2D & 4464 & DDQN & 1457 \\
        5 & N2D & 1054 & N2D & 548 & N2D & 982 \\
        6 & DDQN & 5000 & DDQN & 1559 & DDQN & 20969 \\
        7 & DDQN & 5000 & N2D & 1126 & N2D & 4170 \\
        8 & N2D	& 2237 & N2D & 70 & DDQN & 10411 \\
        9 & DDQN & 5000 & N2D & 2914 & N2D & 1522 \\
        10 & DDQN & 5000 & N2D & 1898 & N2D & 5028 \\
        \hline
        \multicolumn{7}{l}{$^{\mathrm{*}}$DDQN is used by defender and N2D is used by attacker.}
        \end{tabular}
        \label{table_3}
        \end{center}
    \end{table}
    
    
    Table \ref{table_4} shows the control results for the different sets of game 1 without the causative attack active. For set 1 the results were 7 - 3 to three in favour of the defender. On average it took the defender 4,345 turns to win and 1,558 turns for the attacker to win. Set 2 was dominated by the attacker that used the N2D algorithm with 9 wins as the attacker to 1 win for defender On average it took the defender 8,303 amount of turns to win and 3,428 for the attacker to win. Set 3 was dominated by the attacker that used the N2D algorithm with 9 wins as the attacker to 1 win for defender similarly as seen in set 2. The average it took the defender is 27,641 turns to win and of the 9 game runs an average of 3,110 for the attacker.
    
    \begin{table}[htbp]
        \caption{After implementing causative attack in game 2}
        \begin{center}
        \begin{tabular}{|c|c|c|c|c|c|c|}
        \hline
        \textbf{Game}&\multicolumn{6}{|c|}{\textbf{Game sets}} \\
        \cline{2-7} 
        \textbf{Runs} & \multicolumn{2}{|c|}{\textbf{\textit{Set 1}}}& \multicolumn{2}{|c|}{\textbf{\textit{Set 2}}}& \multicolumn{2}{|c|}{\textbf{\textit{Set 3}}} \\
        \cline{2-7}
        \textbf{ } & \textbf{\textit{Agent}}&\textbf{\textit{Turn}}& \textbf{\textit{Agent}}&\textbf{\textit{Turn}}& \textbf{\textit{Agent}}&\textbf{\textit{Turn}}\\
        \hline
        1 & N2D & 40 & N2D & 8992 & DDQN & 27641 \\
        2 & DDQN & 1138 & N2D & 506 & N2D & 2712 \\
        3 & N2D	& 99 & N2D & 1396 & N2D & 8138 \\
        4 & N2D & 4537 & N2D & 3092 & N2D & 162 \\
        5 & DDQN & 4281 & N2D & 7774 & N2D & 80 \\
        6 & DDQN & 5000 & N2D & 5662 & N2D & 8214 \\
        7 & DDQN & 5000 & N2D & 306 & N2D & 2108 \\
        8 & DDQN & 5000 & N2D & 30 & N2D & 670 \\	
        9 & DDQN & 5000 & DDQN & 8303 & N2D & 4310 \\		
        10 & DDQN & 5000 & N2D & 3096 & N2D & 1604 \\		
        
        \hline
        \multicolumn{7}{l}{$^{\mathrm{*}}$DDQN is used by defender and N2D is used by attacker.}
        \end{tabular}
        \label{table_4}
        \end{center}
    \end{table}

\end{enumerate}

\begin{figure}[!tbp]
      \centering
      \begin{minipage}[b]{0.425\textwidth}
        \includegraphics[scale=0.4]{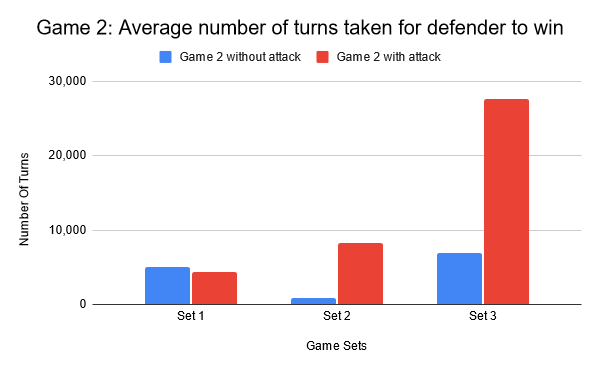}
        \caption{Comparison of average number of turns taken for DDQN as the defender to win.}
        \label{fig:game_2_defender_comp}
      \end{minipage}
      \hfill
      \begin{minipage}[b]{0.425\textwidth}
        \includegraphics[scale=0.4]{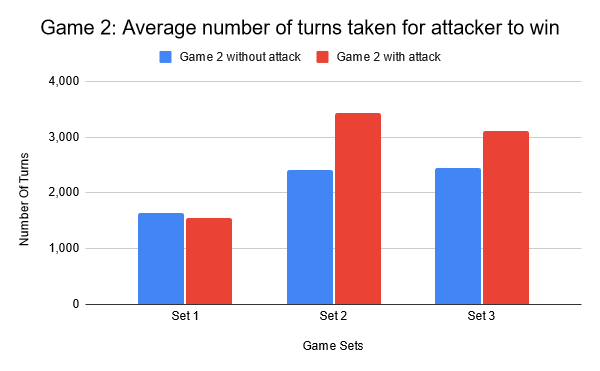}
        \caption{Comparison of average number of turns taken for N2D as the attacker to win.}
        \label{fig:game_2_attacker_comp}
      \end{minipage}
    \end{figure}

The results for each game are analysed as follows:
\begin{enumerate}

    \item \textbf{\textit{Game 1:}}\label{CAG1}
    Figures \ref{fig:game_1_defender_comp} and \ref{fig:game_1_attacker_comp} demonstrate the impact of the attack implementation on the algorithms from both defender and attacker perspectives. In Fig. \ref{fig:game_1_defender_comp}, the results show that the defending agent using the N2D algorithm achieved more wins and improved its turn average after the attack implementation in set 1. In set 2, there was no change in win rates, but the defender's average turn count increased by 32.41\%. For set 3, the defender took longer to isolate the attacker, with a significant increase in turns from 9,534 to 15,593. 
    
    In Fig. \ref{fig:game_1_attacker_comp}, the attacker's average performance improved, but it won fewer games in set 1, indicating efficiency at the cost of consistency. In set 2, there was no change in win rates, but the attacker's average turn count increased by 35.89\%, signifying a loss in performance. However, in set 3, the time taken to capture the flag and win decreased by 32.84\%, indicating a notable improvement. The inclusion of the attack against the defender using NEC2DQN caused a longer time to isolate the attacker and win, decreased performance in set 2, and significant improvement in set 3.

    \begin{figure}[!tbp]
      \centering
      \begin{minipage}[H]{0.425\textwidth}
        \centering
        \includegraphics[scale=0.4]{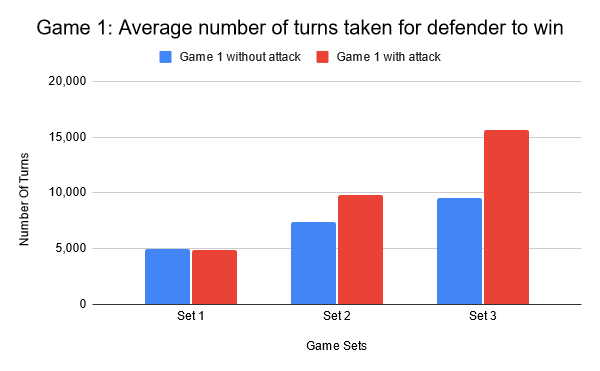}
        \caption{Comparison of average number of turns taken for N2D as the defender to win.}
        \label{fig:game_1_defender_comp}
      \end{minipage}
      \hfill
      \begin{minipage}[H]{0.425\textwidth}
        \includegraphics[scale=0.4]{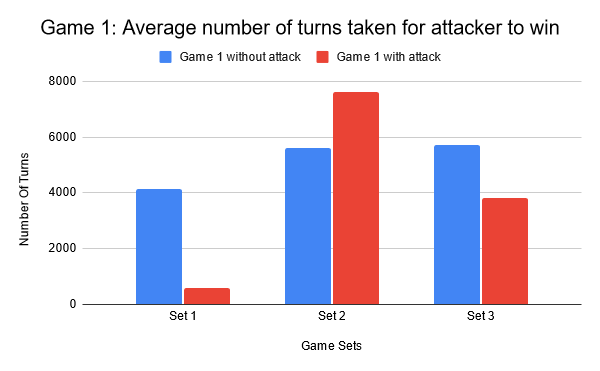}
        \caption{Comparison of average number of turns taken for DDQN as attacker to win.}
        \label{fig:game_1_attacker_comp}
      \end{minipage}
    \end{figure}

    \item \textbf{\textit{Game 2:}}\label{CAG2} Figures            Figures \ref{fig:game_2_defender_comp} and \ref{fig:game_2_attacker_comp} illustrate the impact of the attack implementation on the algorithms from their respective roles. In Fig. \ref{fig:game_2_defender_comp}, the defender's performance is analyzed before and after the attack. For set 1, the defender using the DDQN algorithm won 3 games by isolating the attacker, a notable improvement from previously outlasting the attacker. However, in set 2 and set 3, the defender's wins decreased from 3 to 1 and from 5 to 1, respectively, indicating a clear negative impact from the attack. Examining the average turn count for the lone win in each set becomes irrelevant in this context.
    
    In Fig. \ref{fig:game_2_attacker_comp}, the attacker's perspective is explored before and after the attack implementation. For set 1, the attacking agent using the NEC2DQN algorithm experienced no change in win rate. However, in sets 2 and 3, there was a significant increase in win rates. Despite this improvement, the attacker's average turn count increased by 42.48\% in set 2 and by 26.73\% in set 3. Notably, the impact of the data poisoning attack was greater on the agent using the DDQN algorithm, as it only managed to win one game in both sets, while the attacker secured 9 out of 10 games in sets 2 and 3, albeit with a higher average turn count. 

\end{enumerate}

The experimental results presents two significant implications, regardless of the perspective of the agent. While it may initially seem unfavourable for a defender to struggle in isolating an attacker, the reality offers a silver lining. Prolonged engagements lead to the accumulation of a larger pool of training data, satisfying a core objective of adversarial learning and facilitating the creation of a robust algorithm. The implications of this are discussed in section \ref{conclusion}.

\section{Conclusion and Future Work}\label{conclusion}

This investigation highlights that the DDQN algorithm is more vulnerable to adversarial learning attacks, while NEC2DQN exhibits better resilience. The experiments also show improved engagement and performance of agents in attacking roles and the possibility of training models with adversarial samples during active network engagement. This opens up the potential for an always-online approach without the need for model downtime.

Robust AI model implementation is crucial as attackers constantly strive to break defence mechanisms. In this era of AI and automation, AI systems become the next prime target. Their main vulnerability lies in the learning process, emphasising the importance of developing models robust enough to handle malicious input.

For future work, a more ad hoc network with randomised starting positions and additional defence mechanisms against adversarial attacks will be considered. Potential exploration of partial observability for the attacker and a black box setting will also be examined.

\vspace{12pt}
\begin{IEEEbiographynophoto}{Luke D. Borchjes}
received his Bachelor's Degree (Honours) in Computer Science from the University of the Western Cape. He is currently working on his MSc in Computer Science at the University of the Western Cape. His research
interests are in Deep Reinforcement Learning, Cyber Security and Software Defined Networks.
\end{IEEEbiographynophoto}
\vspace{-1.0cm}

\begin{IEEEbiographynophoto}{Clement N. Nyirenda}
received his PhD in Computational Intelligence from Tokyo Institute of Technology in 2011. His research interests are in Computational Intelligence paradigms such as Fuzzy Logic, Swarm Intelligence, and Artificial Neural Networks and their applications in Communications.
\end{IEEEbiographynophoto}

\vspace{-1.0cm}

\begin{IEEEbiographynophoto}{Louise Leenen}
 Louise Leenen completed her PhD at the University of Wollongong in Australia in 2009. Her research areas are AI Applications in Cybersecurity, Ontology Engineering and Mathematical Modelling.
\end{IEEEbiographynophoto}

\vfill

\end{document}